# Seeing the unseeable

Shep Doeleman
Harvard-Smithsonian Center for Astrophysics

The Event Horizon Telescope, an Earth-sized interferometer, aims to capture an image of a black hole's event horizon to test the theory of general relativity and probe accretion processes.

Black holes are the most mysterious objects predicted by Einstein's theory of general relativity. Yet, despite compelling evidence for their existence, no black hole has been imaged with event-horizon-scale resolution. Supermassive black holes in particular convert the gravitational energy of infalling material into powerful outflows and luminous nuclei that redistribute matter and energy on galactic scales. Horizon-scale access to these central engines of galactic nuclei would enable the most detailed studies of black hole accretion and the formation of relativistic jets, while also providing new tests for the existence of black holes and the validity of general relativity at the boundary of a black hole.

The Event Horizon Telescope (EHT) is an international project specifically designed to capture the first image of a black hole and to time-resolve the dynamics of the associated accretion flow on orbital timescales. To do this, our team links a network of millimetre and submillimetre wavelength observatories around the globe to form an Earth-sized very long baseline interferometry (VLBI) array with the required resolving power and sensitivity. Over the past decade, the EHT has built up the necessary infrastructure by deploying hydrogen maser atomic frequency standards and purpose-built ultrahigh data rate VLBI instrumentation to sites in Chile, Hawaii, Mexico, the South Pole, Arizona and Spain. For the first time this spring, the Atacama Large Millimeter/submillimeter Array (ALMA) facility in Chile joined the array with all of its antennas phased together, operating as a single VLBI station and boosting EHT sensitivity by a factor of ten. Inclusion of ALMA enables the EHT to detect sources within the atmospheric coherence time of several seconds. When observing at 1.3 mm wavelength, the longest baselines in less than 25 microarcseconds.

This resolution is well matched to the two supermassive black holes with the largest apparent event horizons on the sky. Sagittarius A* (Sgr A*), the four million solar mass black hole at the centre of the Milky Way, has a Schwarzschild radius that subtends 10 microarcseconds, and Messier 87 (M87) hosts an approximately six billion solar mass black hole in Virgo A that has a similar angular size. However, extreme light bending near the event horizon forms a 'silhouette' feature bounded by the edge-brightened photon orbit, which is magnified to a size of ~50 microarcseconds. For these sources, observing this silhouette at its expected size would confirm that all the estimated mass is contained within the photon orbit — an extraordinary strengthening of the case for black holes and confirmation of general relativity. The EHT will also search for periodic signals in the VLBI data that correspond to orbits of inhomogeneities in the accretion flow, which are sensitively dependent on black hole spin.

The EHT builds on deep theoretical and technical roots. James Bardeen [1] introduced the idea of the black hole silhouette, and worked out its shape in the general case of a spinning black hole. Subsequent image simulations performed by Jean-Pierre Luminet [2] showed how the silhouette would appear in the presence of accreting material. After the discovery of the compact radio source Sgr A* at the centre of the Milky Way by Bruce Balick and Robert Brown [3], and its identification as the likely emission of a supermassive black hole, a series of VLBI experiments by many investigators pushed to higher and higher frequencies in efforts to resolve the intrinsic structure of the black hole.

The most recent 1.3 mm VLBI results come from an EHT prototype array consisting of stations in Arizona, California and Hawaii. Starting in 2007, observations using this array confirmed the existence of event-horizon-scale structures in Sgr A* [4], detected similarly compact emission at the heart of M87 [5], and revealed ordered and time-variable magnetic fields at the event horizon of Sgr A* [6]. It is this combination of angular resolution and the ability of short millimetre wavelengths to penetrate the accretion flow that provides access to the black hole boundary.

Ten years later, during five days in April 2017, the EHT completed the first in a series of observing campaigns with the potential for producing images. So far, the EHT team has confirmed VLBI detections at all sites using calibrator sources, and processing of the data is ongoing. It is too early to tell what the results will be, but the EHT Collaboration is aiming to carry out yearly observations in coordination with planned enhancements of the array. These improvements include increased bandwidth, the addition of new stations, and extension to observing wavelengths of 0.87 mm, where angular resolution increases. It may be that multiple observing runs, higher sensitivity and higher resolution will be needed to image the anticipated strong gravity features. We should know soon.

The EHT Collaboration combines an international group of institutes and facilities with support from multiple science funding agencies (www.eventhorizontelescope.org).